\documentclass[%
superscriptaddress,
twocolumn,
amsmath,amssymb,
 aps,
prx,
]{revtex4-2}
\usepackage[dvipsnames]{xcolor}
\usepackage{graphicx}% Include figure files
\usepackage{dcolumn}% Align table columns on decimal point
\usepackage{bm}% bold math
\usepackage[caption=false]{subfig}
\usepackage[colorlinks]{hyperref}
\hypersetup{%
	plainpages=true,
	breaklinks=true,%not default in dvips mode, so we must specify
	hypertexnames=false,%not ideal, but needed when pagenums duplicate (`i' vs. `1')
	pageanchor=true,
	colorlinks=true,
	linkcolor={blue},
	citecolor={red},
	urlcolor={blue},
	anchorcolor={black}
}% add hypertext capabilities
\usepackage{braket,slashed}
\usepackage{tikz}
\usepackage{tikzit}
\usetikzlibrary{quantikz}

\usepackage{algorithm}
\usepackage{algpseudocode}

\newcommand{\setT}{\Delta}
\newcommand{\setone}{\delta_1}
\newcommand{\settwo}{\delta_2}

\newcommand{\bA}{\bar{A}}

\newenvironment{diagram}
{
\begin{tikzpicture}[baseline = (X.base),every node/.style={scale=0.7},scale=.55]
}
{
\end{tikzpicture}
}

\newcommand{\mpshort}
{\begin{diagram}
\draw (0.5,1.5) -- (1,1.5); 
\draw[rounded corners] (1,2) rectangle (2,1);
\draw (1.5,1.5) node (X) {$A$};
\draw (2,1.5) -- (3,1.5); 
\draw[rounded corners] (3,2) rectangle (4,1);
\draw (3.5,1.5) node {$A$};
\draw (4,1.5) -- (5,1.5);
\draw[rounded corners] (5,2) rectangle (6,1);
\draw (5.5,1.5) node {$A$};
\draw (6,1.5) -- (7.5,1.5); 
\draw (1.5,1) -- (1.5,.5); 
\draw (3.5,1) -- (3.5,.5); 
\draw (5.5,1) -- (5.5,.5);
\end{diagram}}

\newcommand{\overlapshort}
{
\begin{diagram}
\draw (0.5,1.5) -- (1,1.5); \draw (0.5,-1.5) -- (1,-1.5); 
\draw[rounded corners] (1,2) rectangle (2,1);
\draw[rounded corners] (1,-1) rectangle (2,-2);
\draw (1.5,1.5) node {$A_1$};
\draw (1.5,-1.5) node {$\bar{A}_2$};
\draw (2,1.5) -- (3,1.5); \draw (2,-1.5) -- (3,-1.5);
\draw[rounded corners] (3,2) rectangle (4,1);
\draw[rounded corners] (3,-1) rectangle (4,-2);
\draw (3.5,1.5) node {$A_1$};
\draw (3.5,-1.5) node {$\bar{A}_2$};
\draw (4,1.5) -- (5,1.5); \draw (4,-1.5) -- (5,-1.5);
\draw[rounded corners] (5,2) rectangle (6,1);
\draw[rounded corners] (5,-1) rectangle (6,-2);
\draw (5.5,1.5) node {$A_1$};
\draw (5.5,-1.5) node {$\bar{A}_2$};
\draw (6,1.5) -- (6.5,1.5); \draw (6,-1.5) -- (6.5,-1.5);
\draw (1.5,1) -- (1.5,-1); 
\draw (3.5,1) -- (3.5,-1);
\draw (5.5,1) -- (5.5,-1);
\end{diagram}
}

\newcommand{\transfer}{
\begin{diagram}
\draw[rounded corners] (1,1.5) rectangle (2,.5);
\draw[rounded corners] (1,-1.5) rectangle (2,-.5);
\draw (1.5,1) node {$A_1$}; \draw (1.5,-1) node (X) {$\bar{A}_2$};
\draw (1.5,.5) -- (1.5,-.5); 
\draw (1,1) -- (.5,1); \draw (2,1) -- (2.5,1);
\draw (1,-1) -- (.5,-1); \draw (2,-1) -- (2.5,-1);
\draw (1.5,0) node (X) {$\phantom{X}$};
\end{diagram}
}

\newcommand{\expectation}{
\begin{diagram}
\draw (1,-1.5) edge[out=180,in=180] (1,1.5);
\draw[rounded corners] (1,2) rectangle (2,1);
\draw[rounded corners] (1,-1) rectangle (2,-2);
\draw (1.5,0) circle (.5);
\draw (1.5,0) node {$O_{0}$};
\draw (1.5,1.5) node {$A_L$};
\draw (1.5,-1.5) node {$\bar{A}_L$};
\draw (1.5,1) -- (1.5,.5);
\draw (1.5,-1) -- (1.5,-.5);
\draw (2,1.5) -- (3,1.5); 
\draw (2,-1.5) -- (3,-1.5);
\draw[rounded corners] (3,2) rectangle (4,1);
\draw[rounded corners] (3,-1) rectangle (4,-2);
\draw (3.5,0) circle (.5);
\draw (3.5,0) node {$O_{1}$};
\draw (3.5,1.5) node {$A_L$};
\draw (3.5,-1.5) node {$\bar{A}_L$};
\draw (3.5,1) -- (3.5,.5);
\draw (3.5,-1) -- (3.5,-.5);
\draw (4,1.5) -- (4.5,1.5); 
\draw (4,-1.5) -- (4.5,-1.5);
\draw (5.,0)  node[auto=false]{\ldots};
\draw (6,1.5) -- (5.5,1.5); 
\draw (6,-1.5) -- (5.5,-1.5);
\draw[rounded corners] (6,2) rectangle (7,1);
\draw[rounded corners] (6,-1) rectangle (7,-2);
\draw (6.5,1.5) node {$A_L$};
\draw (6.5,-1.5) node {$\bar{A}_L$};
\draw (6.5,1) -- (6.5,.5);
\draw (6.5,-1) -- (6.5,-.5);
\draw (6.5,0) circle (.5);
\draw (6.5,0) node {$O_{n-1}$};
\draw (7,1.5) -- (8,1.5); \draw (7,-1.5) -- (8,-1.5);
\draw[rounded corners] (8,2) rectangle (9,1);
\draw[rounded corners] (8,-1) rectangle (9,-2);
\draw (8.5,1.5) node {$A_C$};
\draw (8.5,-1.5) node {$\bar{A_C}$};
\draw (8.5,0) circle (.5);
\draw (8.5,0) node {$O_n$};
\draw (8.5,1) -- (8.5,.5);
\draw (8.5,-1) -- (8.5,-.5);
\draw (8.5,0) circle (.5);
\draw (9,+1.5) edge[out=0,in=0] (9,-1.5);
\end{diagram}
}

\begin{document}

\usetikzlibrary{quantikz}

\title{Quantum kernels to learn the phases of quantum matter}% Force line breaks with \\
\author{Teresa Sancho-Lorente}
\affiliation {Instituto de Nanociencia y Materiales de Aragón (INMA)
  CSIC-Universidad de Zaragoza, Zaragoza 50009,
  Spain}

\author{Juan Román-Roche}
\affiliation {Instituto de Nanociencia y Materiales de Aragón (INMA)
  CSIC-Universidad de Zaragoza, Zaragoza 50009,
  Spain}
  
\author{David Zueco}
\affiliation {Instituto de Nanociencia y Materiales de Aragón (INMA)
  CSIC-Universidad de Zaragoza, Zaragoza 50009,
  Spain}
\date{\today}% It is always \today, today,
             %  but any date may be explicitly specified

\begin{abstract}
Classical machine learning has succeeded in the prediction of both classical and quantum phases of matter. Notably, kernel methods stand out for their ability to provide interpretable results, relating the learning process with the physical order parameter explicitly. Here, we exploit quantum kernels instead. They are naturally related to the \emph{fidelity} and thus it is possible to interpret the learning process with the help of quantum information tools. In particular, we use a support vector machine (with a quantum kernel) to predict and characterize second order quantum phase transitions.
We explain and understand the process of learning when the fidelity per site (rather than the fidelity) is used.
The general theory is tested in the Ising chain in transverse field. We show that for small-sized systems, the algorithm gives accurate results, even when trained away from criticality. Besides, for larger sizes we confirm the success of the technique by extracting the correct critical exponent $\nu$. 
Finally, we present two algorithms, one based on fidelity and one based on  the fidelity per site, to classify the phases of matter in a quantum processor.
\end{abstract}

%\keywords{Suggested keywords}%Use showkeys class option if keyword
                              %display desired
\maketitle

%\tableofcontents

%%%%%%%%%%%%%%%%%%%%%%%%%%%%%%%%%%%%%
%%%%%%%%%%%%%%%%%%%%%%%%%%%%%%%%%%%%%
%%%%%%%%%%%%%%%%%%%%%%%%%%%%%%%%%%%%%
%%%%%%%%%%%%%%%%%%%%%%%%%%%%%%%%%%%%%
\section{Introduction}
It is suggestive to merge quantum computing  and machine learning (ML), looking for their constructive combination
in hopes of increasing the number of problems that can be solved in the near future.
Both are disruptive technologies that cross the boundaries of current computational capabilities.
%
%In few words, and in a probably too naive way, ML  deals with   large sets of  data.
%The hope is that a quantum computer can handle them when  a classical one cannot. 
%
%Things are not so easy, as always.
%
\emph{Classical} ML has found several applications for optimizing tasks in quantum information processing.  Some examples are quantum state tomography \cite{Torlai2018,Shahnawaz2021} quantum gate optimization \cite{Niu2019, An2019, Zhang2019},  ground state estimation \cite{Carleo2017} among others.  See the reviews \cite{Carleo2019, Carrasquilla2020}.
\emph{Quantum} machine learning (QML), instead, seeks to extend the algorithms of classical ML to be run in a quantum computer \cite{Schuld2014, Biamonte2017, PerdomoOrtiz2018}.  
An incomplete list of  reported examples  are the quantum versions of neural networks \cite{Altaisky2001},  principal component analysis \cite{Lloyd2014}, classification \cite{Schuld2020, PerezSalinas2020, dutta2021, Bartkiewicz2020, belis2021} or  support vector machines \cite{Rebentrost2014}. 
The key question here is whether they have any kind of advantage over their classical counterparts \cite{Tang2019, Tang2019a, Arrazola2020,Huang2021}.
\\
Both in classical and quantum ML, input data is encoded in $M$ vectors ${\boldsymbol x}_j$, $j=1,\ldots,M$ of dimension $d$.
They are processed in different ways, depending on the chosen algorithm and/or type of learning.
In this work we are interested in \emph{kernel} methods \cite{schoelkopf2002}.
Here, the learning is based on the kernel, which is the inner product of these vectors $ K_{ij} = \boldsymbol x_i \cdot \boldsymbol x_j$ or, more generally, the inner product defined in a \emph{feature space}.  
The latter is given by a non-linear map $\boldsymbol x_j \to \boldsymbol \Phi (\boldsymbol x_j)$.   Thus, $K_{ij} =  \boldsymbol \Phi (\boldsymbol x_i) \cdot  \boldsymbol \Phi (\boldsymbol x_j)$.
To understand this, we can think of a classification task where the data is split into two classes. 
The feature map should clump data belonging to the same class while dispersing data from different classes, so that the resulting distribution is  separable. The kernel defines distances on the feature space, on which  classification takes place. 
\\
In QML, data is loaded in a quantum computer, thus, the first step is \emph {encoding} it onto a quantum state: 
\begin{equation}
\label{qfeature}
{\bf x}_j \to
|\psi ({\bf x}_j) \rangle =U_\theta ({\bf x}_j) | 0 \rangle \; .
\end{equation}
Here, $| 0 \rangle$ is the initial state \footnote{
 Encoding may be generalised to a map onto density matrices ${\bf x}_j \to \varrho ({\bf x}_j)$} and $\theta$ a set of parameters that, eventually, can be optimized.  
Thus, the quantum kernel is given by \cite{Rebentrost2014, Schuld2019,LaRose2020, schuld2021}:
\begin{equation}
\label{qK}
    K_{ij}^{(\theta)} = | \langle \psi_\theta ({\bf x}_j) |\psi_\theta ({\bf x}_{j^\prime}) \rangle |
    = 
    | \langle 0 | U_\theta^\dagger ({\bf x}_i) \; U_\theta ({\bf x}_j)  | 0 \rangle | \; .
\end{equation}

%Therefore, it can be obtained by evolving with $U_\theta^\dagger ({\bf x}_i)  U_\theta ({\bf x}_j)$ and measuring in the computational basis \cite{Havlek2019, Agarwal2021}.   
%

%This completes the algorithm to be run:
%\\
%\begin{equation}
%\label{qcircuit}
%\begin{quantikz}
%\lstick{$\ket{0}$}    
%&  \gate{{U_\theta}^\dagger ({\bf x}_i)}  \qwbundle[alternate]{} 
%& \gate{{U_\theta}({\bf x}_j)^{\color{white} \dagger}} %\qwbundle[alternate]{} 
%& \meter{} \qwbundle[alternate]{}
%\end{quantikz}
%\begin{quantikz}
%\lstick{$\ket{0}$} 
%& \gate[wires=2][1cm]{U_\theta^\dagger ({\bf x}_i) }
%& \qw & \gate[wires=2][1cm]{U_\theta({\bf x}_j)}
%& \qw  & \meter{}  \\
%\lstick{$\ket{0}$} 
% & & \qw  & \qw  & \qw & \meter{}
%\end{quantikz}
%\end{equation}
%This circuit is among the simplest, others are possible based on fidelity estimations \cite{Agarwal2021}. 

Mapping \eqref{qfeature}  offers   a quantum advantage if the quantum circuit is difficult to simulate on a classical computer and provides a better performance than classical maps.
It is not trivial to obtain instances of quantum advantage.
A heuristic approach, implementing entangled maps that are shown to be classically hard, has been followed in \cite{Havlek2019, peters2021}.
From these seminal works, attempts to (rigorously) determine under what conditions q-kernels are superior have been discussed \cite{huang2021a, wang2021}. 
Finally, a quantum speed-up has been shown for the task of classifying integers according to the, so-called,  discrete
logarithm problem \cite{liu2020}.
This is quite a formal problem so the challenge posed in the first section of this letter persists, the identification of tasks of \emph{practical} use where QML is advantageous.
\\

One possible shortcut to achieving the goal is to consider quantum data.
The idea is simple: generating the data already requires a quantum computer and the step of loading classical data onto a quantum RAM is skipped.
The task we propose is classifying the phases of matter.
For classical models, classical ML techniques have been discussed with both kernel methods \cite{Ponte2017, Liu2019, Giannetti2019} and beyond \cite{Wang2016, Carrasquilla2017, Hu2017, Wetzel2017, Beach2018, Schfer2019, MendesSantos2021, maskara2021}.
For quantum models, neural networks trained with different observables   \cite{Biamonte2017,vanNieuwenburg2017,Chng2017, Broecker2017}, among other ML techniques \cite{Che2020, Lidiak2020, huang2021b} or, even experimentally,  with a quantum simulator   \cite{Bohrdt2019}  have been used to classify phases in strongly correlated electron systems.  
Here, we propose to use quantum kernels   as \eqref{qK} \cite{Banchi2021, wu2021provable}.  Then, the classification is done with a support vector machine (SVM).
Borrowing from quantum information, we argue that, by employing the fidelity (or related measures) as a kernel, the classification can be interpreted to extract the phase boundary \cite{Zanardi2006, zhou2008a, Zhou2008b, zhou2007}. 
We show that the machine predicts the critical point and that it is capable of learning the critical exponents.
 Remarkably, it does so despite being trained with samples far from the critical point.
Here, by way of illustration, we use the one-dimensional Ising model in a transverse field.
This is an exactly solvable model, but our arguments are pretty general.
In fact, we present two algorithms to classify the quantum states of matter in the general scenario where the ground states are computed in a quantum processor. 
To be concrete, we discuss the use of a variational quantum eigensolver to find the ground state  \cite{Peruzzo2014, Kandala2017}.

We overview the rest of the paper here.
In the next section \ref{sec:fidel}, the theory of fidelity-based characterization of QPTs is summarised following the seminal works of Zanardi.  
Then, in Sect. \ref{sect:svms} we sketch the idea behind of SVMs and the Kernel trick.  Quantum Kernels are introduced.  
Importantly, in sect. \ref{sec:theo} we discuss the process of learning and how the machine learns to characterize a second order QPT.
It explains the results of  the rest of the paper.
Our general theory is tested in the quantum Ising model in one dimension in sect. \ref{sec:ising}.
We also present two algorithms to implement the ideas of this paper in a quantum computer, see sect. \ref{sec:alg}.
The paper ends, as usual, with the conclusions in \ref{sec:conc}.
Some identities for quantum states, used throughout the paper, can be found in the appendix \ref{app:A}.
%

%%

%%%%%%%%%%%%%%%%%%%%%%%%%%%%
%%%%%%%%%%%%%%%%%%%%%%%%%%%%
%%%%%%%%%%%%%%%%%%%%%%%%%%%%
%%%%%%%%%%%%%%%%%%%%%%%%%%%%
\section{Quantum Phase Transitions and Fidelities}
\label{sec:fidel}

%Generally speaking a phase transition  occurs when the state of the system 

%In this letter we are interested in quantum phase transitions.
%
Consider a Hamiltonian $H(J)$ such that at  $J=J_c$ the system undergoes a quantum phase transition (QPT).
Whether first, second order or topological, the QPT can be studied from the \emph{distinguishability} between  ground states.
Following the original idea of Zanardi and Paunkovi{\'{c}} a measure of this distinguishability is the fidelity between two ground states,
\begin{equation}
\label{Fdef}
  F (J, J^\prime ) := | \langle \psi_0 (J)| \psi_0 (J^\prime) \rangle| \; .
\end{equation}
In a nutshell, and considering
$F(J, J+ \epsilon)$  with $\epsilon$ small enough, only at the critical point (or close enough) $F$ is expected to deviate from $1$. 
Thus, an abrupt change in the fidelity signals criticality \cite{Zanardi2006, Cozzini2007} (See \cite{Gu2010} for a review).
Following this idea,  Zhou and coworkers argue in terms of renormalization  \cite{zhou2008a, Zhou2008b, zhou2007, zhou2008c}.  
This is specially useful for continuous QPTs, occuring at the  thermodynamic limit.
The rest of this letter assumes this type of QPT.
\\

In what follows, we find it convenient to discuss general expressions for the fidelity between two quantum states using the MPS formalism.  
This allows to anticipate the $N$-dependence of the fidelity and to introduce a new distance measure between ground states that will be used throughout this work.
In the case of a translational invariant lattice, with local dimension $l$, the ground state can be written in its Matrix Product State (MPS) form \cite{Orus2014},
%\begin{equation}
$ |    \psi_0 (J) \rangle
=
{\rm Tr} [ A_{i_1} ... A_{i_N}] |i_1, ..., i_N \rangle 
$.
%\end{equation}
Here, $\{A_{i_j}\}$ are  $D \times D$ matrices that depend on the local quantum number $i_j=1,..., l $. $D$ is the bond dimension, which  is related to the amount of entanglement contained in the state. 
Using the MPS representation, the fidelity takes the convenient form \cite{Cozzini2007}, see also our App. \ref{app:A}:
\begin{equation}
\label{Flambda}
    F (J, J^\prime) = \sum_{k=1}^{D^2} \lambda_k (J, J^\prime) ^N 
    \stackrel{N\gg 1}{\cong}\lambda_1(J, J^\prime) ^N \; .
\end{equation}
Here, $\lambda_k$ are the eigenvalues of the transfer matrix $E(J, J^\prime) = \sum_{i=1}^l A_i(J) \otimes A_i(J^\prime)$. 
The state is normalized so $|\lambda_k| \leq 1$.  If we denote $\lambda_1$ the largest (in absolute value) of these eigenvalues, the second equality is obvious and motivates the definition of the \emph{fidelity per site}
\begin{equation}
\label{lambdadef}
  \log [\lambda (J, J^\prime)] :=  \; \log F[(J, J^\prime)]/N   \; .
\end{equation}
Importantly, 
it inherits the properties of $F$, being thus a distance measure fully characterizing the QPT.
Besides, it has an important advantage (over $F$). %At sufficiently large $N$, $\lambda \cong \lambda_1$ it is size independent, thus 
In \eqref{Flambda} the orthogonality catastrophe is explicit. 
For $ N $ large enough, two ground states with $ J \neq J^\prime $ have exponentially small fidelity, regardless of whether or not they belong to the same phase.
This alerts to the failure of $ F $ as a distance measure to resolve the transition.
Using \eqref{Flambda} and \eqref{lambdadef} we note that $ \lambda_1 = \lim_ {N \to \infty} \lambda $, \emph {i.e.} $\lambda_1$ is a scale factor independent of size. 
The use of the fidelity per site as a measure prevents the orthogonality catastrophe.

%

%%%%%%%%%%%%%%%%%%%%%%%%%%%%%%%%%%%%%%%%%%%%%%%%%%
%%%%%%%%%%%%%%%%%%%%%%%%%%%%%%%%%%%%%%%%%%%%%%%%%%
%%%%%%%%%%%%%%%%%%%%%%%%%%%%%%%%%%%%%%%%%%%%%%%%%%
%%%%%%%%%%%%%%%%%%%%%%%%%%%%%%%%%%%%%%%%%%%%%%%%%%
\section{Phase classification and SVMs}
\label{sect:svms}
Identifying the phases in quantum many body systems  can be formulated as a classification task. 
To simplify the discussion, let us assume that the system has two phases separated at $J=J_c$.
Classifying means assigning a label  $y_J = \pm 1$  to every  ground state $| \psi_0 (J) \rangle$ depending on  $J \gtrless  J_c$ respectively.
In this work,  the training data are the ground states themselves, \emph{i.e}
${\bf x}_{j}= |\psi_0 (J_j)\rangle$.
In other words, we choose a set $\{J_j\}$ and  compute the corresponding g.s, presumably in a quantum computer.   
This is the training set, used in a supervised learning algorithm for classifying the data.
In this paper we use a support vector machine (SVM).
This  algorithm finds the hyperplane that optimally splits the data in two, given a training set \cite{schoelkopf2002}. 
It turns out that the hyperplane is found by minimizing  a Lagrangian:
$ L ({\bf \alpha}) =  \sum \alpha_j - \frac{1}{2} \sum \alpha_i \alpha_j y_i y_j K({\bf x}_i, {\bf x}_j)$, that depends on the kernel,
with constraints $\sum \alpha_j y_j=0$. The $\alpha_j$'s are the Lagrange multipliers found in the Lagrangian minimization. Only a subset of them will be non-null, attending to Eq. \eqref{distance}. These determine the classification, i.e. define the separating hyperplane. They are termed support vectors (SV). Then, given a ground state $| \psi_0 (J)\rangle$,  its signed ``distance'' to the hyperplane is given by
\begin{equation}
\label{distance}
d (J)=\sum_{j=1}^M \alpha_j y_{j} K(J, J_j) + b \,.
\end{equation}
where $b$ is the offset parameter (we use here the standard notation).   It is given by,
\begin{equation}
    b=\frac{1}{M} \sum_{i,j} \alpha_i y_i K(x_i, x_j) - y_j \; .
\end{equation}

If the optimization is successful, the separating hyperplane will lie at the phase boundary between the two phases and new data points will be classified attending to the sign of their ``distance'' to the hyperplane.  
\\
The crux of the matter is the kernel matrix. 
The better it measures the similarity between the data points the more it facilitates classification.
Based on our fidelity discussion, we are going to consider two kernels that, as argued above, can measure the distance between quantum states and thus, are useful to discriminate different phases.
Using Eqs. \eqref{qK}, \eqref{Fdef} and \eqref{lambdadef} we can define:
\begin{align}
\label{KF}
    K^{(F)} (J_i, J_j) & := F (J_i, J_j) \; , 
    \\
    \label{Kl}
 K^{(\lambda)} (J_i, J_j) & := \lambda (J_i, J_j) \; .
\end{align}
\\
We expect that the fidelity-based kernel $K^{(F)}$ will fail for a sufficiently large system size $N$ (orthogonality catastrophe). In this case, the kernel is ``overfitted''. See also Ref. \cite{Banchi2021, shaydulin2021importance}.  Therefore, it seems that it cannot fully characterize QPTs at the thermodynamic limit.  However, below we show that $K^{(\lambda)}$ learns the QPT.  This will be complemented (in Sect. \ref{sec:ising}) with numerical simulations where we will also test the failure of $K^{(F)}$ as the system size grows.

\subsection{What does the SVM learn?}
\label{sec:theo}
\begin{figure} [h]
    \includegraphics[width = 1.\columnwidth]{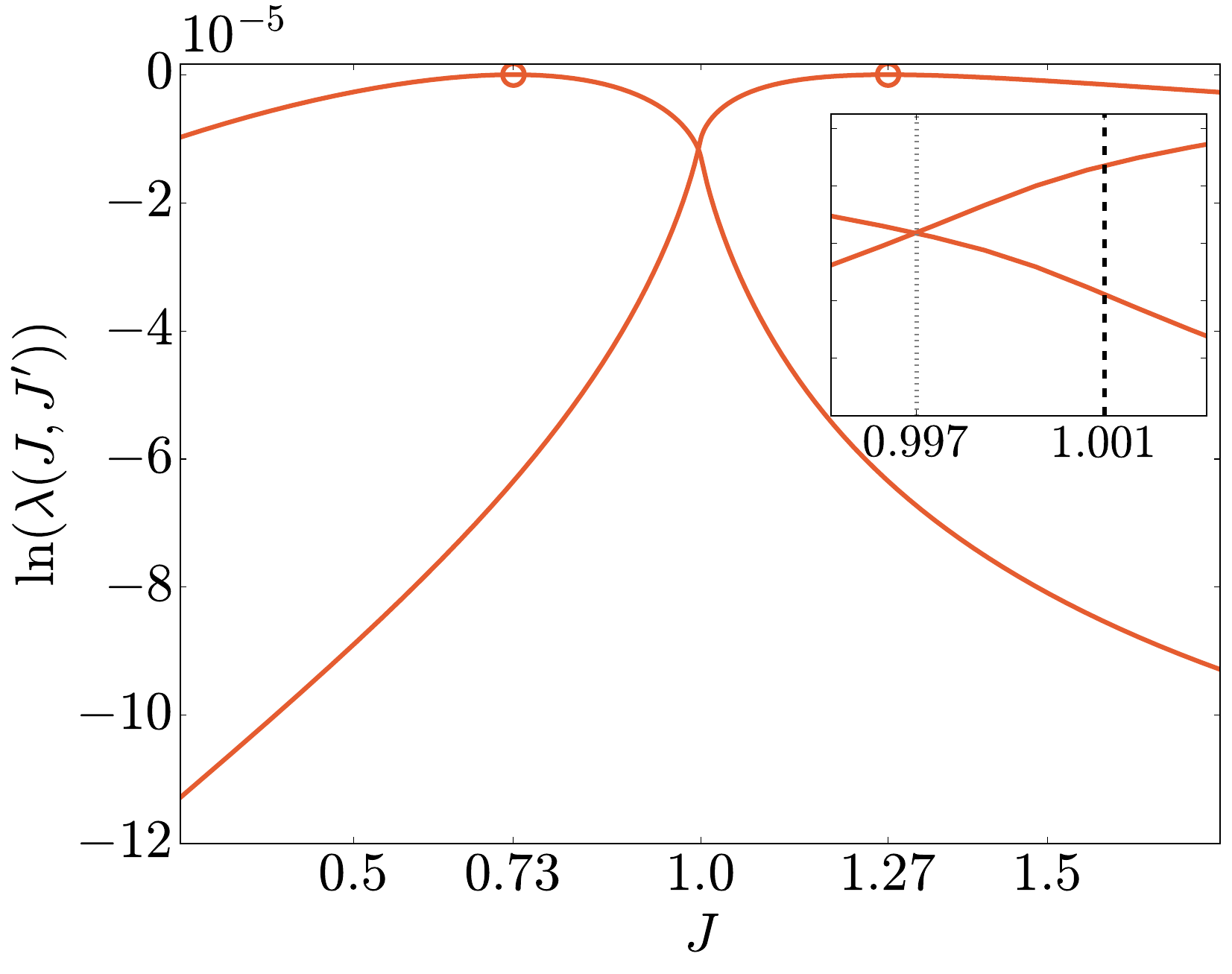}
    \caption{{\bf Critical point from the SVM.}  $\lambda (J_\pm, J)$ in the one dimensional quantum Ising model, see Eq. \eqref{eqn:ising}, with $J_{-}=0.73$ and $J_{+}=1.27$ for $N=1000$ spins. $J_{\pm}$ are represented by the open circles.  {\bf Inset}:  Zoom of the intersection. The gray dotted line points out the intersection between the two curves, \emph{i.e.} the $J$ at which $d=0$, and so the boundary predicted by the SVM, $\widetilde J_c(N)$.  The maximuum of the derivative $\partial_x \lambda (J_+, x)$ is marked with the black dashed line, $J_c(N)$.}
    \label{fig:argument}
\end{figure}

In this section we argue that, under reasonable conditions (to be explained),  \emph{the SVM learns the critical $J_c$ in a  2nd order QPT using  the Kernel} $K^{(\lambda)} (J_i, J_j) := \lambda (J_i, J_j) $,  Eq. \eqref{Kl}.
Before, some remarks are necessary.
We assume that the data is separable. As one sweeps (in values of $J$) across the phase transition, for every new $J$, the ground state $\ket{\phi(J)}$ is increasingly further (fidelity closer to zero) from any given ground state in the other phase.  In other words, there is a direction in the Hilbert space endowed with distance $\lambda (J_i, J_j)$ such that the projection of the ground states $\ket{\phi(J)}$ along this direction is a strictly monotonic function of $J$. This is (by definition) what happens in a phase transition when we look at it in parameter space: phases are separated by a clear boundary, i.e. the critical point. What we are assuming is that this intuitive property applies as well to the Hilbert space when the distance $\lambda (J_i, J_j)$ is used. This view is supported by the results in \cite{Zanardi2006, Zhou2008b, Gu2010}.
A separable set allows us to train the SVM with a hard-margin, which will have consequences for the result of the learning process. Mainly that, barring a failure of the learning process due to a deficient kernel, we will find only two SVs, one on each side of the separating hyperplane. 
Let us denote by $|\psi (J_\pm) \rangle$ the SVs on the right and left side of the transition (respectively). Besides, the ground state  located at the separating hyperplane, $d(\widetilde J_c) =0$, is $|\psi (\widetilde J_c) \rangle$. Consequently, if $\widetilde J_c = J_c$,  the SVM has learnt the critical point.

Let us first prove that for QPTs and using $K^{(\lambda)}$,  $b=0$ in  \eqref{distance}.
We use that only  SVs have a non-null multiplier,  $\alpha_j \neq 0$,  and that they are equidistant to the separating hyperplane,  $\lambda(J_+,  \widetilde J_c ) = \lambda(J_-,  \widetilde J_c )$.  
Therefore,
\begin{equation*}
    b= \frac{\lambda (J_+,  \widetilde J_c )}{2} \left(\alpha_+ - \alpha_- \right) =0 \; .
\end{equation*}
Here, we have used the  constraint: $\sum \alpha_j y_j=0$ which implies $\alpha_+ = \alpha_-$.
If $b=0$, then $d(\widetilde J_c) \propto \lambda (J_+, \widetilde J_c) -  \lambda (J_+, \widetilde J_c) =0$.  
%%In other words, the hyperplane is not biased and occurs  at the intersection of the two curves $\lambda (J_\pm, x)$.  In fact,  this is how  $\widetilde %%J_c$ is  obtained.
%
To show how the intersection of the two curves $\lambda (J_\pm, J)$ relates to the QPT, we plot Fig. \ref{fig:argument}. It shows a generic situation of how two fidelity curves intersect. 
It is a calculation using the quantum Ising model, to be discussed below, but other models with a 2nd order QPT show the same phenomenology.
Starting from $\lambda(J_+, J_+)=1$ the fidelity remains close to one.   
As $J \to J_c$, see Sect. \ref{sec:fidel}, there is a non-analyticity in $\lambda (J_+, J)$, with a sudden increase in the slope. This is the signature of the transition. The same occurs for $\lambda (J_-, J)$.
\ For our purposes here, it means the intersection of the two curves occurs in the vicinity of $J_c$.
The non-analyticity corresponds to the point where  $\partial_x \lambda (J_\pm, x)$ is maximum.
In fact, a way of finding QPTs is by looking at this maximum, \emph{i.e.} defining $J_c(N)$ as the point where the derivative is maximum, then $J_c = \lim_{N\to \infty} J_c(N)$.
%
%Putting it all together, we can always write
%\begin{equation}
%\begin{split}
%    \lambda (J_-, \widetilde J_c(N)) &=   \lambda (J_-,  J_c(N)) \\ 
%    &+ (\widetilde J_c(N)-J_c(N)) \partial_{x=J_c(N)} \lambda (J_-, x)  
%    \\
%    &+ \mathcal O ( \widetilde J_c(N)-J_c(N))^3 \; .
%    \label{taylor}
%\end{split}
%\end{equation}
%
Since the slope $\partial_{x=J_c(N)} \lambda (J_-, x)$ grows with $N$, we find  that the larger the $N$ the closer the intersection moves towards $J_c(N)$, i.e. the closer  $\widetilde J_c(N)$ to $J_c(N)$.
In that case, the SVM learns the transition point as predicted by the fidelity theory.
\\

%%%%%%%%%%%%%%%%%%%%%%%%%%%%%%%%%%%%%%%%%%%%%%%%%%
%%%%%%%%%%%%%%%%%%%%%%%%%%%%%%%%%%%%%%%%%%%%%%%%%%
%%%%%%%%%%%%%%%%%%%%%%%%%%%%%%%%%%%%%%%%%%%%%%%%%%
%%%%%%%%%%%%%%%%%%%%%%%%%%%%%%%%%%%%%%%%%%%%%%%%%%
\section{Application: the quantum Ising model}
\label{sec:ising}
%%%%%%%%%%%%%%%%%%%%%%%%%%%%%%%%
%%%%%%%%%%%%%%%%%%%%%%%%%%%%%%%
\begin{figure*}
    \centering
    \includegraphics[width = 1.\textwidth]{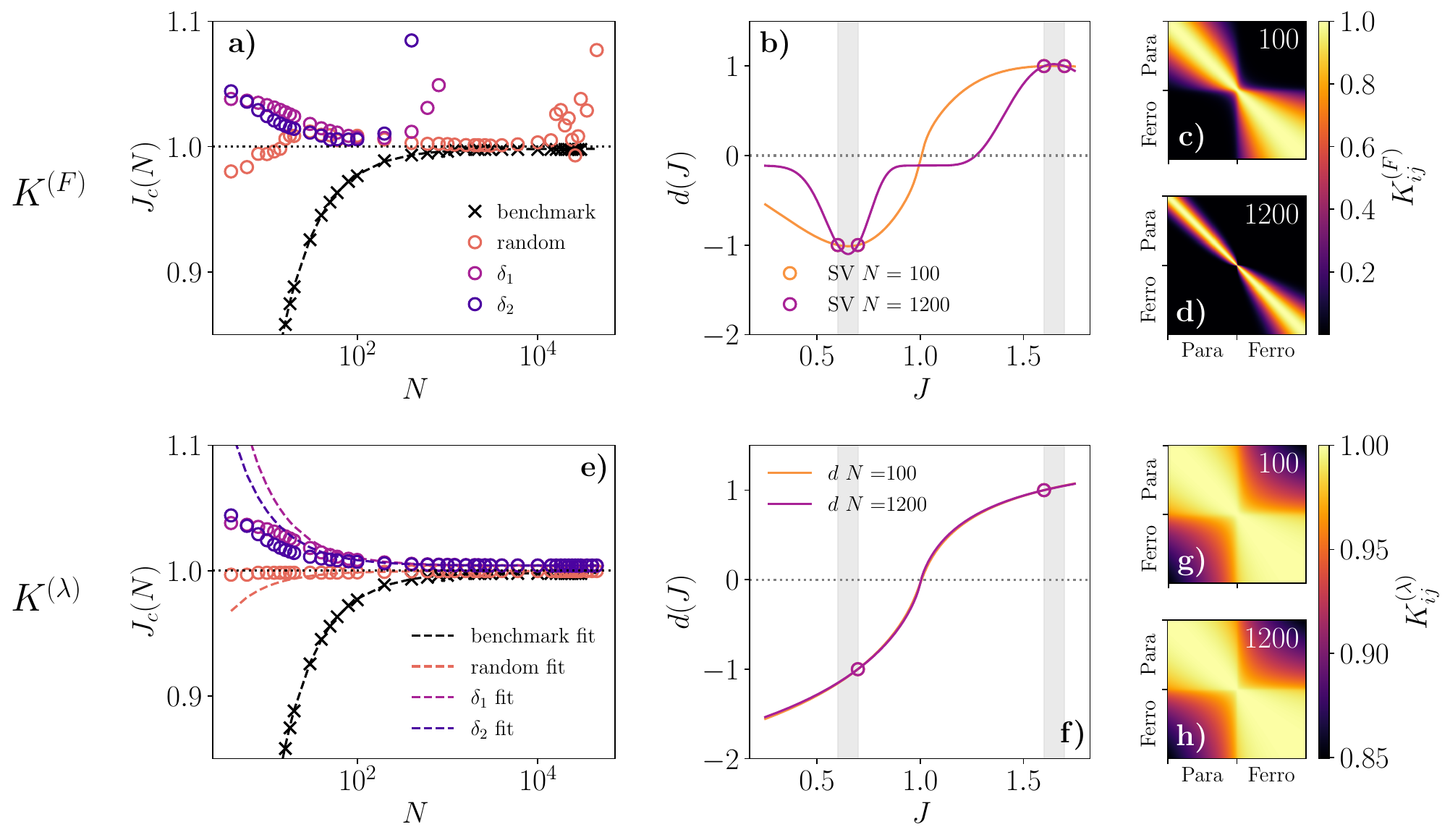}
    \caption{{\bf Quantum kernels learning phases of matter.}   {\bf a)} $J_c (N)$  using the kernel $K^{(F)}$ for the three trainings discussed in this work. Namely,  by intervals taking  points failing in the $\setT$-subsets
$\setone = [0.8, 0.9] \cup [1.2, 1.3]$,  
 $\settwo= [0.6, 0.7] \cup [1.6, 1.7]$ and   taking  $M=133$ points randomly distributed over $\setT$ (random). The set $\setT$ are 
  $1000$ ground states taking $J$s equally spaced in $\setT= [0.25, 1.75]$.   Open circles are SVM results using the class {\tt sklearn.svm.SVC} from Sklearn \cite{scikit-learn}.  The crosses stand for a finite size scaling analysis following \cite{Zhou2008b}.  The corresponding dashed lines are  fittings (see table \ref{table}). 
{\bf b)} Distance function, Eq. \eqref{distance}, for two  sizes $N$.   The open circles are the SVs and the shaded (gray) zones mark the training interval, $\settwo$.  
{\bf c)} and {\bf d)}  Contour plots of the matrix $K^{(F)}_{ij}$ where $i,j$ run on the total  $\setT$.  The labels mark the (theoretical) phases of the model for $J_i$ ($J_j$).  Lower row, panels {\bf e), f), g) and h)} are the counterparts but using the kernel $K^{(\lambda)}$.}
    \label{fig:svm}
\end{figure*}
%%%%%%%%%%%%%%%%%%%%%%%%%%%%%%
%%%%%%%%%%%%%%%%%%%%%%%%%%%%%%
We consider the  one dimensional quantum Ising model in a transverse field with Hamiltonian ($\hbar =1$ through this letter)
\begin{equation}
    \label{eqn:ising}
    H (J) =\sum_{i=1}^N {\sigma}_j^z-J\sum_{i=1}^N {\sigma}_j^x {\sigma}_{j+1}^x \; .
\end{equation} 
$\sigma_j^\alpha$ are the Pauli matrices acting on the $j$-lattice site.
The lattice size  is $N$.  Periodic boundary bonditions (PBC) are considered.
This Ising model has a second order  phase transition  occurring  at $ J_c =1 (-1)$ in the $N\to \infty$ limit. For $J_c >1 (J_c<-1)$  the $\mathbb{Z}_2$ (parity) symmetry is spontaneously broken and the g.s is ferromagnetically (antiferromagnetically) ordered. 
W.l.o.g. we  fix our attention in the paramagnetic-ferromagnetic transition occuring at $J_c=1$.
%and it is  exactly solvable.
%
\\
For our interest here, the QPT has been discussed in terms of the ground
state fidelity $F(J, J^\prime)$ and the fidelity per site 
$\lambda (J, J^\prime)$, Eq. \eqref{Fdef} and \eqref{lambdadef} \cite{Zanardi2006, Zhou2008b, Gu2010}.
Thus, it is an ideal testbed for our proposal.
Besides,
it is exactly solvable via the Jordan Wigner (JW) transformation, so an explicit formula for $F$ between two ground states can be found:
\begin{equation}
\label{FJW}
F(J_i, J_j) 
%= \langle \psi_0(J_j) \mid \psi_0 (J_{j^{\prime}})\rangle
=  \prod_{k}  \left| \cos \left(
\theta_{k} (J_i)-\theta_ k (J_{j})
\right) \right| \; .
\end{equation}
Here,
%$|\psi_0 (J_{j})\rangle$ is the ground state of \eqref{eqn:ising} for a given coupling strength $J_{j}$.  
\begin{equation}
    \cos (2 \theta_{k} (J_j) ) = \frac{1 +2 J_j \cos k}{\sqrt{1 + 4 J_j \cos k + 4 J_j^2}} \,.
\end{equation}
Considering even $N$ and PBC, $k = (2 n -1) \pi /N$ with $n=1, ..., N/2$.  
\\

With \eqref{FJW} at hand,   kernels \eqref{KF} and \eqref{Kl} can be computed.
In this letter,  the data sets are obtained by 
 taking $1000$ $J$'s equally spaced in the range $\setT = [0.25, 1.75]$.
 %so the Kernel matrices $K^{(F)} (J_i, J_j)$ and $K^{(\lambda)} (J_i, J_j)$   
 %
For the training set, we explore three possibilities.   
Two consist on training with $J$'s belonging only to specific intervals.  
From $\setT$, we take the points falling in the subsets
$\setone = [0.8, 0.9] \cup [1.2, 1.3]$ and 
 $\settwo= [0.6, 0.7] \cup [1.6, 1.7]$.
Notice that they are non-symmetric  (around $J_c=1$) in order to challenge our algorithm abilities in the classifying task.
 Besides,  and importantly enough, they are far away from the critical value.
The third training set is formed by taking $M=133$ points randomly distributed over the whole set $\setT$ \footnote{The number $M=133$ is fixed to match the number of points entering in both $\setone$ and $\settwo$}.
\\
Our main results are summarized in the different panels of Fig. \ref{fig:svm}. 
The top row of panels contains results for the fidelity kernel $K^{(F)}$ \eqref{KF}.  They are compared to the results from the $\lambda$-kernel $K^{(\lambda)}$ \eqref{Kl} (bottom row). 
In  panel \ref{fig:svm} a) we plot the predicted $\widetilde J_c(N)$ , using the kernel $K^{(F)}$.
This is obtained from the distance function \eqref{distance} after interpolating the point that fulfills  $d (\widetilde J_c(N)) \stackrel{!}{=} 0$. 
We do it for the three training sets (empty circles) as a function of  $N$.
For small sizes, the tendency  is expected.  The estimation of $\widetilde J_c(N)$ improves with the system size $N$.
Going to larger sizes, $K^{(F)}$ fails. 
To understand why, in Fig. \ref{fig:svm} b) we plot the distance to the separating hyperplane \eqref{distance}.  We only show the $\setone$-interval training.
 The other training sets behave similarly and can be found in \cite{SM}.  
For small $N$ the  distance is a smooth function, which confirms that the SVM is able to generalize well to the test data and provides a good estimation of $\widetilde J_c(N)$.
Increasing $N$, the distance function flattens in the critical region, hindering the extrapolation of $\widetilde J_c(N)$.
The explanation for this failure, as we have anticipated, is
the  orthogonality catastrophe. In Fig. \ref{fig:svm} c) and d) we show further proof of this, we plot
the kernel matrix $K^{(F)} (J_i, J_j)$ for the whole set $\setT$.  For small sizes ($N=100$) there is a block structure, marking the ability to distinguish the two phases.  On the other hand, for $N=1200$, all entries are close to zero (bar the diagonal), i.e. any two states are orthogonal.  This is in accordance with the fact that the training using the interval $\setone$ breaks down earlier as it is the furthest away from the critical point, while random training has points closest to the transition point and breaks down last.  The corresponding Kernels trained with intervals can be found in \cite{SM}.
\\
One way to get around the catastrophe is by using $K^{(\lambda)}$ instead. 
For small sizes, both kernels give comparable predictions.
However, as shown in panel \ref{fig:svm} e) the prediction always improves with $N$, approaching $\widetilde J_c (N) \to J_c \cong 1$ in the limit   $N\to \infty$, see table \ref{table}.  This confirms what was said in Sect. \ref{sec:theo}.
In panel  \ref{fig:svm} f) we plot the distance,  confirming  convergence in the thermodynamic limit [Cf. panel \ref{fig:svm}b)].
Finally, the kernel matrix shows a marked block diagonal structure at any lattice size. See Fig. \ref{fig:svm} g) and h) and compare them to their counterparts  d) and e) respectively. \\
To benchmark the SVM predictions, after our last discussion in \ref{sec:theo} and following \cite{Zhou2008b} we can define $J_c(N)$ as the point at which the function $\partial_J \log \lambda (J, J^\prime = 1.75)$ is maximal. For a fair comparison, we do it on the same set of $J$'s ($\setT$) as the SVM training. The SVM works better at small lattice sizes.
%
%A \emph{tentative} explanation is that trainings use ground state far away from criticality.
A \emph{tentative} explanation is that our SVM training sets $\delta_1$ and $\delta_2$ are comprised of ground states far from criticality (less so for the random set), whereas the benchmark uses the full dataset  $\setT$, which includes ground states close to the transition.
This is relevant because far from criticality the correlation length is finite, and not-too-large systems seem to be sufficient for learning.
These are good news for medium-sized quantum processors.

%%%%%%%%%%%%%%%%%%%%%%%%%%%%%%%%%%%%%%%%%%%%
%%%%%%%%%%%%%%%% table %%%%%%%%%%%%%%%%%%%%%
%%%%%%%%%%%%%%%%%%%%%%%%%%%%%%%%%%%%%%%%%%%%
\begin{table}
\begin{ruledtabular}
\begin{tabular}{lll}
&  $J_c$ & $\nu$ \\ 
\hline \\[-1.8ex] 
$\lambda$  as \cite{Zhou2008b} & 0.99827(12) & 0.966(17)\\
SVM  ($\setone$) & 1.00414 $(*)$ & 0.974(60)\\
SVM  ($\settwo$) & 1.00404 $(*)$ & 1.003(75)\\
SVM  (random) & 0.99975 $(*)$ & 0.966(46)\\
\end{tabular}
\end{ruledtabular}
\caption{\label{table} $J_c$ and $\nu$ critical exponent results.  The numbers are obtained by fitting the points in Fig. \ref{fig:svm} e) as a function of $N$ to the function $J_c(N)= J_c +a \, N^{-\nu}$ with fitting parameters $a$, $J_c$ and $\nu$. $J_c(N)$ are given by the SVM algorithm.  They are compared to the procedure developed in \cite{Zhou2008b} based on the fidelity per site, $\lambda$ (see main text).  The $(*)$  means that the error given by the fitting is smaller than $10^{-8}$.   Other error sources, as the number of training data or the $\setT$ discretization limit the accuracy. }
\end{table}

In addition, for the SVM to characterize the QPT, we must check if it is capable of learning the critical exponents.
For thermal transitions, the critical exponents are learnt  when the distance \eqref{distance} can be related to the order parameter \cite{Giannetti2019, Ponte2017}, such that the distance inherits the scaling exponents of the latter. 
In our case, the distance to the hyperplane is a linear combination of the fidelity per site.
Thus, we expect to have  the same finite scaling as $\lambda$, from which the corresponding critical parameters can be extracted. 
This is plotted in  Fig. \ref{fig:svm} e). 
Dashed lines are the best fittings to the scaling formula,
\begin{equation}
\label{JcN}
|J_c - J_c (N)| \sim N^{-1/\nu} \; . 
\end{equation}
The fitted $\nu$ are summarized in table \ref{table}. For the Ising model, Hamiltonian \eqref{eqn:ising}, $\nu=1$. Thus, the SVM with  $K^{(\lambda)}$ is able to learn the critical exponent.

%%%%%%%%%%%%%%%%%%%%%%%%%%%%%%%%%%%%%%%%%%%%%%%%%%
%%%%%%%%%%%%%%%%%%%%%%%%%%%%%%%%%%%%%%%%%%%%%%%%%%
%%%%%%%%%%%%%%%%%%%%%%%%%%%%%%%%%%%%%%%%%%%%%%%%%%
%%%%%%%%%%%%%%%%%%%%%%%%%%%%%%%%%%%%%%%%%%%%%%%%%%

\section{Quantum Algorithms}
\label{sec:alg}
\begin{figure*}
    \centering
    \includegraphics[width = 1.\textwidth]{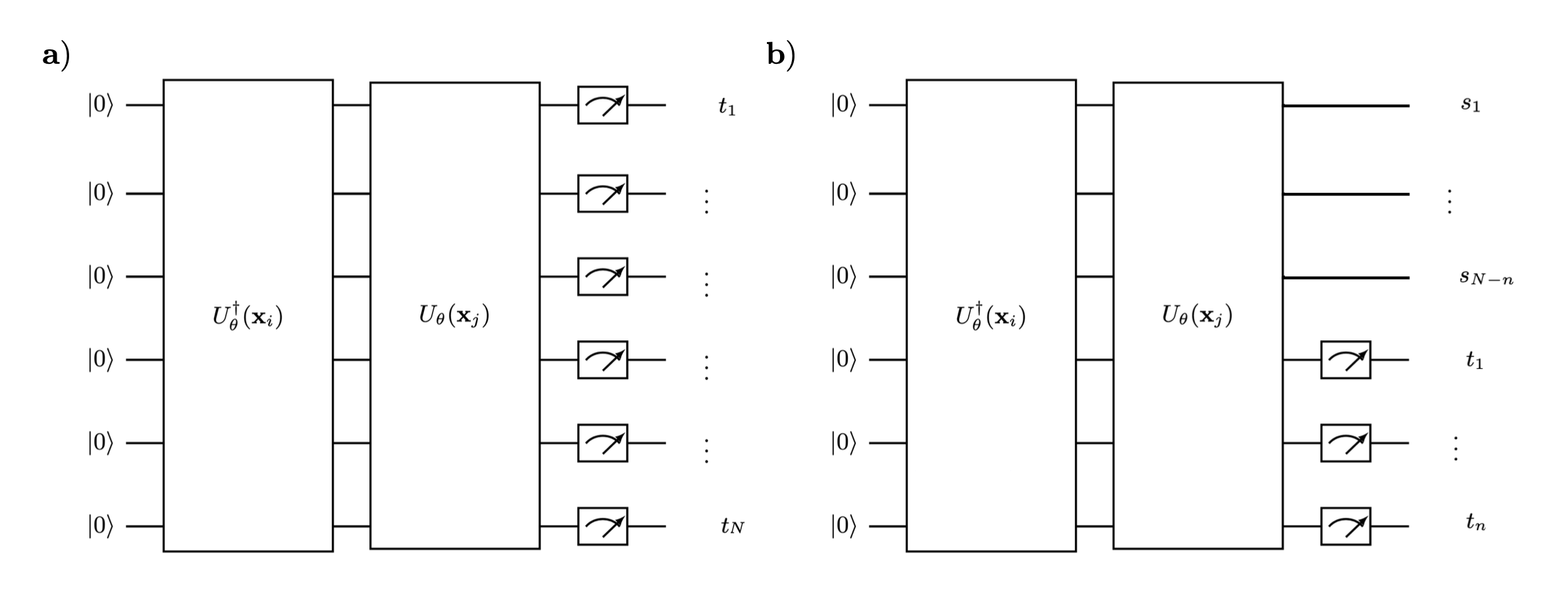}
    \caption{{\bf Classification algorithms} {\bf a)} Trial quantum circuit constructed to estimate the fidelity between quantum stated. This measurements are the key of the classification of the phases in the system of study.  {\bf b)} Our proposal of quantum algorithm to estimate the fidelity per site in the chain, minimizing the scaling errors. }
    \label{fig:circuits}
\end{figure*}

The Ising model allowed us to demonstrate the usefulness of quantum kernels and their performance in the classification of quantum phases.  
Our arguments  are both interpretable and based on the wave-function, thus they are exportable to other Hamiltonians, Cf. Sect. \ref{sec:theo}.
In particular, to those where the ground states can be obtained within a quantum processor.
For those cases, we introduce here two algorithms  for computing $K^{(F)} (J_i, J_j)$ and $K^{(\lambda)} (J_i, J_j)$ respectively. (See Fig. \ref{fig:circuits})

While $K^{(F)} (J_i, J_j)$ fails for a sufficiently large system, it gives a very good estimate of $J_c$ for medium sizes, which is the realistic situation within the NISQ era.
Algorithm \ref{alg:1}:
\begin{algorithm}[H]
  \caption{ Classification using $K^{(F)}$ }
  \label{alg:1}
   \begin{algorithmic}[1]
   \Require Training set $\{J_j, y_j\}_{j=1}^M$, \emph{i.e.} parameter values of the parameterized Hamiltonian  $H(J)$ along with the labels of the corresponding phase,  $y_j=\pm 1$.
   \State Compute the corresponding ground states.  Here, we are thinking of a VQA algorithm, where the circuit depends on some variational parameters $\theta_j$:
   $$| \psi (J_j) \rangle = U_{\theta_j} | 0 \rangle \; .$$
   $U_{\theta_j}$ is the quantum circuit.
   \State Store the classical parameters $\theta_j$ (in a classical memory).
   %\State 
   \State Prepare the circuit
   $$ U_{\theta_i}^\dagger \, U_{\theta_j} |0^N \rangle \; ( \equiv | \psi_{J_i, J_j} \rangle)$$
   \If {$i=j$}
   $$K_{ii}^{(F)}=1$$
   \Else $\;$ Measure all bits for the state $| \psi_{J_i, J_j} \rangle$ in the computational basis.  The frequency, $p_{0^N}$  of the all-zero outcome  corresponds to the state overlap, \emph{i.e.} the Kernel entrance
   $$(K^{(F)}_{ij})^2 = p_{0^N} $$ 
   see fig. \ref{fig:svm} a).
   \EndIf 
   \State Use SVM (hard margin).
   \end{algorithmic}
\end{algorithm}
Notice that the depth of the circuit to calculate any kernel entrance $K_{ij}$ is the sum of the depths to obtain
    the corresponding $|\psi(J_i)\rangle$ and $|\psi(J_j)\rangle$.  The complexity scales as $\mathcal O (\epsilon^{-2} M^4)$.  Here, $\epsilon$ is the largest sampling error $\epsilon \sim \mathcal{O} (R^{-1/2})$.  $R$ is the number of shots to estimate $p_{0^N}$ \cite{Havlek2019}.  

In principle, the most demanding part is obtaining  ground states (step 1).  
It is hard even for a quantum computer.  This task is within  the QMA complexity class \cite{Kempe2004}, roughly speaking the NP-complete analogue for quantum computers. 
Nevertheless, 
quantum computers can be better than classical methods such as density functional theory \cite{Argaman2000}, density normalization group \cite{Schollwock2005}, tensor networks \cite{Orus2014}, quantum montecarlo \cite{Troyer2009} or even ML-inspired techniques \cite{Carleo2017}, in certain cases. 
See a recent discussion  in \cite{schiffer2021}.  In particular, heuristic quantum algorithms such as adiabatic  \cite{Albash2018} or varational ones \cite{farhi2014} can be efficient for some Hamiltonians.
This has been shown, for example,  for  non critical spins systems \cite{BravoPrieto2020}.  
This justifies the combined use of SVM and quantum computing. We have found that classification is successful even if trained away from criticality.

\subsection{Fidelity per site algorithm}

A complete characterization of a QPT requires a scaling analysis of the fidelity per site, so an algorithm to compute this quantity must be devised.
An option would be to compute the $N$th root of each element $ K_{ij}^{(\lambda)}= [K_{ij}^{(F)}]^{1/N}$.
It does not work.  
The inevitable error in $K_{ij}^{(F)}$ explodes (with $N$) when performing the $N$th root.  

This can be fixed by modifying algorithm \ref{alg:1} as follows.
After step $3$ there,  if only  $n<N$  qubits are measured  [Cf. figures \ref{fig:svm} a) and b)] the frequency, $p_{0^n}$ of  all-zero outcome
corresponds to 
\begin{equation}
\label{fn}
    f_n^2 := \langle O_n \rangle _{J, J^\prime} \sim  (\lambda_1^n)^2
\end{equation}
Here, $O_n = | 0^n \rangle \langle 0^n |$ and $\langle \; \; \rangle_{J, J^\prime}$ is the expectation using the state $| \psi_{J_i, J_j} \rangle)$ [Cf. algorithm \ref{alg:1}].
For the second relation we use  \eqref{expectationmps}.   The latter follows from further identities    summarized in App. \ref{app:A}.
Furthermore, using \eqref{Flambda} and \eqref{lambdadef}, $\lambda_1 \to \lambda$ in the thermodynamic limit.
Therefore, 
$\lambda$ can be inferred  by repeating the protocol for different $N$s and fitting the value of the scaling parameter.   This completes our second  algorithm:
\begin{algorithm}[H]
  \caption{Classification using $K^{(\lambda)}$ }
  \label{alg:2}
   \begin{algorithmic}[1]
   \Require  Same as in algorithm \ref{alg:1}.
   \State Steps 1, 2 and 3 are as in algorithm \ref{alg:1}
   \If {$i=j$}
   $$K^{(\lambda)}_{ii}=1$$
   \Else 
   \State Initialise $n$ such that $ 1 \ll n \ll N$.
   \For {$l =0, 1, ..., R^{\prime}$}
    
    Measure $n$ qubits $(t_1, t_2... t_n)$, see fig. \ref{fig:svm} b).
    
    Obtain $f_n^2$, Eq. \eqref{fn}.
   
    $n=n+l$
    
    \EndFor
  
    \State Fit $K_{ij}^{(\lambda)} :=  \lambda_{ij}  \sim (f_n)^{1/n}$.  
   \EndIf
   \State Use SVM (hard margin).
   \end{algorithmic}
\end{algorithm}

Like in algorithm \ref{alg:1}
the complexity scales as $\mathcal O (\epsilon^{-2} M^4)$.  Now,  $\epsilon \sim \mathcal{O} ((R R^\prime)^{-1/2})$.

%%%%%%%%%%%%%%%%%%%%%%%%%%%%%%%%%%%%%%%%%%%%%%%%%%
%%%%%%%%%%%%%%%%%%%%%%%%%%%%%%%%%%%%%%%%%%%%%%%%%%
%%%%%%%%%%%%%%%%%%%%%%%%%%%%%%%%%%%%%%%%%%%%%%%%%%
%%%%%%%%%%%%%%%%%%%%%%%%%%%%%%%%%%%%%%%%%%%%%%%%%%
\section{Conclusions}
\label{sec:conc}
In this work we have discussed the capabilities of quantum kernels to  characterize QPTs.  
For second order QPTs, we have shown that, using the fidelity per site as kernel, a SVM  characterizes a QPT: both the transition point and critical exponents.
Let us emphasize that  the  kernels rely on the fidelity, and thus solely on the wave function. Accordingly,  the classification does not need any previous knowledge of the order parameter or the symmetries of the Hamiltonian. 

The theory has been applied in the quantum Ising model. This is an exactly solvable model, for non-solvable models we have presented two quantum algorithms, one based on the Fidelity and the other on the Fidelity per site.  
In both of them, the most expensive part is obtaining the ground states. Since the SVM training can done out of criticality, we argued that this  alleviates this task.

We believe that this work shows a fruitful synergy between quantum information processing (in this work for obtaining the quantum data, \emph{i.e.} the ground states)  and ML (here, for learning the phases of matter) in a classically hard problem \cite{Cirac2012, Georgescu2014}. 

\acknowledgements
The authors acknowledge enlightening discussions with Javier Molina-Vilaplana, Juanjo García-Ripoll and Eduardo Sánchez-Burillo. Funding from the EU (COST Action 15128 MOLSPIN, QUANTERA SUMO and FET-OPEN Grant 862893 FATMOLS), the Spanish MICINN, the Gobierno de Arag\'on (Grant E09-17R Q-MAD) and the CSIC 
Quantum Technologies Platform PTI-001.

%%%%%%%%%%%%%%%%%%%%%%%%%%%%%%%%
%%%%%%%%%%%%%%%%%%%%%%%%%%%%%%%%
%%%%%%%%%%%%%%%%%%%%%%%%%%%%%%%%
%%%%%%%%%%%%%%%%%%%%%%%%%%%%%%%%
\appendix

\section{Uniform Matrix Product States identities}
\label{app:A}

Let us follow \cite{Vanderstraeten2019}.  Any translational invariant  state can be written as,
\begin{align}
\nonumber
| \psi \rangle = &\sum_{\{s\}} \boldsymbol{v}_{L}^{\dagger}\left[\prod_{m \in \mathbb{Z}} A^{s_{m}}\right] \boldsymbol{v}_{R}|\{s\}\rangle 
\\
    & %\tikzfig{figure0}[scale=2]
    = \cdots \mpshort \cdots
    \label{psimps}
\end{align}
The boundary vectors  $\boldsymbol{v}_{L}$ and $\boldsymbol{v}_{R}$ must be irrelevant for larger sizes.  In the thermodynamical limit, physical states cannot depend on the boundary conditions.  Those
where the boundary conditions do matter have  measure zero in the space of all possible MPS tensors.

The fidelity can be written:
\begin{align}
\nonumber
\braket{\Psi(\bA_2)|\Psi(A_1)} & = 
\dots 
\overlapshort
\dots 
\\ \nonumber
& =
\lim_{N\to \infty}
\left (
\transfer
\right )^N
\\ 
& = \lim_{N\to \infty} E^N
= \lim_{N\to \infty}  \lambda_1^N
\end{align}
Here, $\bar A$ is the complex conjugate of $A$, $E= A_1 \otimes \bar A_2$ is the, so called, transfer matrix. $\lambda_1$ is the leading $E$-eigenvalue. This formula is used in the main text, Sect. \ref{sec:fidel}.

The state is uniquely defined by the tensor $A$.  
The opposite is not true.
Different tensors can yield the same state.  
In fact, the \emph{gauge transform} $A \to X A X^{-1}$, leaves the state \eqref{psimps} invariant.
This being said,  it is convenient
to introduce left and right canonical forms $A_L = L A L^{-1}$ ($A_R = R A R^{-1}$) such that   identities are :
\begin{subequations}
\begin{equation}
\label{AL}
\begin{diagram}
\draw (1,-1.5) edge[out=180,in=180] (1,1.5);
\draw[rounded corners] (1,2) rectangle (2,1);
\draw[rounded corners] (1,-1) rectangle (2,-2);
\draw (1.5,1) -- (1.5,-1);
\draw (1.5,1.5) node {$A_L$};
\draw (1.5,-1.5) node {$\bar{A}_L$};
\draw (2,1.5) -- (2.5,1.5); 
\draw (2,-1.5) -- (2.5,-1.5);
\end{diagram} = 
\begin{diagram}
\draw (1,-1.5) edge[out=180,in=180] (1,1.5);
\end{diagram}\;, 
\end{equation}
%%%%%%
\begin{equation}
\begin{diagram}
\label{AR}
\draw (0.5,1.5) -- (1,1.5); \draw (0.5,-1.5) -- (1,-1.5);
\draw[rounded corners] (1,2) rectangle (2,1);
\draw[rounded corners] (1,-1) rectangle (2,-2);
\draw (1.5,1) -- (1.5,-1);
\draw (1.5,1.5) node {$A_R$};
\draw (1.5,-1.5) node {$\bar{A}_R$};
\draw (2,-1.5) edge[out=0,in=0] (2,1.5);
\end{diagram} = 
\begin{diagram}
\draw (1,-1.5) edge[out=0,in=0] (1,1.5);
\end{diagram}\;.
\end{equation}
\end{subequations}

Canonical forms are useful for computing observables. 
For our purposes, it is sufficient consider observables of the form
\begin{equation}
    O = O_0 \otimes ... \otimes O_{n}
\end{equation}
where $O_0$ acts on a single site that w.l.o.g we can label as $0$-site, then  $0_{i}$ acts on the site $i$ respect to this $0$-site.  Using the canonical forms \eqref{AL} and \eqref{AR} the expectation value can be computed as:
\begin{equation}
\label{expectationmps}
\langle O \rangle = 
    \expectation
\end{equation}
Here, $A_C = L^{-1} A R$.  This formula is used in algorithm  \ref{alg:2}.
%%%%%%%%%%%%%%%%%%%%%%%%%%%%%%%%
%%%%%%%%%%%%%%%%%%%%%%%%%%%%%%%%
%%%%%%%%%%%%%%%%%%%%%%%%%%%%%%%%
%%%%%%%%%%%%%%%%%%%%%%%%%%%%%%%%
%\bibliography{../main}% Produces the bibliography via BibTeX.
\bibliography{main}
\end{document}